\documentclass[10pt,twoside,twocolumn,british,english,aps,superscriptaddress,reprint]{revtex4-1}
\usepackage{lmodern}
\usepackage{lmodern}
\usepackage[LGR,T1]{fontenc}
\usepackage[latin9]{inputenc}
\usepackage{color}
\usepackage{babel}
\usepackage{array}
\usepackage{textcomp}
\usepackage{amsmath}
\usepackage{amssymb}
\usepackage{graphicx}
\usepackage[unicode=true,pdfusetitle,
 bookmarks=true,bookmarksnumbered=true,bookmarksopen=false,
 breaklinks=true,pdfborder={0 0 0},pdfborderstyle={},backref=false,colorlinks=true]
 {hyperref}
\hypersetup{
 linkcolor=red, citecolor=blue,  urlcolor=blue}

\makeatletter

\DeclareRobustCommand{\greektext}{%
  \fontencoding{LGR}\selectfont\def\encodingdefault{LGR}}
\DeclareRobustCommand{\textgreek}[1]{\leavevmode{\greektext #1}}
\ProvideTextCommand{\~}{LGR}[1]{\char126#1}

\providecommand{\tabularnewline}{\\}

\usepackage{babel}
\usepackage{natbib}

\makeatother

\begin{document}
\title{Structural, thermodynamic, and local probe investigations of a honeycomb
material Ag$_{3}$LiMn$_{2}$O$_{6}$}
\author{R. Kumar}
\affiliation{Department of Physics, Indian Institute of Technology Bombay, Powai,
Mumbai 400076, India}
\author{Tusharkanti Dey}
\affiliation{Experimental Physics VI, Center for Electronic Correlations and Magnetism,
University of Augsburg, D-86159 Augsburg, Germany}
\author{P. M. Ette}
\affiliation{Central Electrochemical Research Institute-Madras Unit, CSIR-Madras
Complex, Taramani, Chennai 600113, India}
\author{K. Ramesha}
\affiliation{Central Electrochemical Research Institute-Madras Unit, CSIR-Madras
Complex, Taramani, Chennai 600113, India}
\author{A. Chakraborty}
\affiliation{School of Physical Sciences, Indian Association for the Cultivation
of Science, Jadavpur, Kolkata 700 032, India}
\author{I. Dasgupta}
\affiliation{School of Physical Sciences, Indian Association for the Cultivation
of Science, Jadavpur, Kolkata 700 032, India}
\author{R. Eremina}
\affiliation{Kazan (Volga Region) Federal University, Kremlevskaya st., 18, Kazan,
420008, Russia}
\affiliation{Kazan E. K. Zavoisky Physical-Technical Institute (KPhTI) of the Kazan
Scientific Center of the Russian Academy of Sciences, Sibirsky tract,
10/7, Kazan, 420029, Russia }
\author{Sándor Tóth}
\affiliation{Laboratory for Neutron Scattering and Imaging, Paul Scherrer Institute
(PSI), CH-5232 Villigen, Switzerland}
\author{A. Shahee}
\affiliation{Department of Physics, Indian Institute of Technology Bombay, Powai,
Mumbai 400076, India}
\author{S. Kundu}
\affiliation{Department of Physics, Indian Institute of Technology Bombay, Powai,
Mumbai 400076, India}
\author{M. Prinz-Zwick}
\affiliation{Experimental Physics V, Center for Electronic Correlations and Magnetism,
University of Augsburg, D-86159 Augsburg, Germany}
\author{A.A. Gippius}
\affiliation{Department of Physics, M.V. Lomonosov Moscow State University, 199991
Moscow, Russia}
\affiliation{P.N. Lebedev Physics Institute of Russian Academy of Science, 199991
Moscow, Russia}
\author{H. A. Krug von Nidda}
\affiliation{Experimental Physics V, Center for Electronic Correlations and Magnetism,
University of Augsburg, D-86159 Augsburg, Germany}
\author{N. B{ü}ttgen }
\affiliation{Experimental Physics V, Center for Electronic Correlations and Magnetism,
University of Augsburg, D-86159 Augsburg, Germany}
\author{P. Gegenwart }
\affiliation{Experimental Physics VI, Center for Electronic Correlations and Magnetism,
University of Augsburg, D-86159 Augsburg, Germany}
\author{A.V. Mahajan}
\email [Corresponding author: ] {mahajan@phy.iitb.ac.in}

\affiliation{Department of Physics, Indian Institute of Technology Bombay, Powai,
Mumbai 400076, India}
\date{\today}
\begin{abstract}
Here we present the structural and magnetic properties of a new honeycomb
material Ag$_{3}$LiMn$_{2}$O$_{6}$. The system Ag{[}Li$_{1/3}$Mn$_{2/3}${]}O$_{2}$
belongs to a quaternary 3R-delafossite family and crystallizes in
a monoclinic symmetry with space group $C\,2/m$ and the magnetic
Mn$^{4+}$($S=3/2$) ions form a honeycomb network in the $ab$-plane.
An anomaly around 50\,K and the presence of antiferromagnetic (AFM)
coupling (Curie-Weiss temperature $\theta_{CW}\sim-51$\,K) were
inferred from our magnetic susceptibility data. The magnetic specific
heat clearly manifests the onset of magnetic ordering in the vicinity
of 48\,K and the recovered magnetic entropy, above the ordering temperature,
falls short of the expected value, implying the presence of short-range
magnetic correlations. An asymmetric Bragg peak (characteristic of
two dimensional order), seen in neutron diffraction, gains intensity
even above the ordering temperature, thus showing the existence of
short-range spin correlations. Our electron spin resonance ESR experiments
corroborate the bulk magnetic data. Additionally, the (ESR) line broadening
on approaching the ordering temperature $T_{{\rm N}}$ could be described
in terms of a Berezinski-Kosterlitz-Thouless (BKT) scenario with $T_{{\rm KT}}=40(1)$~K.\textcolor{black}{{}
$^{7}$Li NMR line-shift probed as a function of temperature tracks
the static susceptibility (}\textit{\textcolor{black}{K}}\textcolor{black}{$_{iso}$)
of magnetically coupled Mn$^{4+}$ ions. The $^{7}$Li spin-lattice
relaxation rate (1/$T$$_{1}$) exhibits a sharp decrease below about
50\,K. A critical divergence is absent at the ordering temperature
perhaps because of the filtering out of the antiferromagnetic fluctuations
at the Li site, }\textit{\textcolor{black}{i.e.}}\textcolor{black}{,
at the centers of the hexagons in the honeycomb network. Combining
our bulk and local probe measurements, we establish the presence of
an ordered ground state for the honeycomb system }Ag$_{3}$LiMn$_{2}$O$_{6}$\textcolor{black}{.
}Our\textbf{ }\textit{ab~initio}\textbf{ }electronic structure calculations
suggest that in the $ab$-plane, the nearest neighbor (NN) exchange
interaction is strong and AFM, while the next NN and the third NN
exchange interactions are FM and AFM respectively. The interplanar
exchange interaction is found to be relatively small. In the absence
of any frustration the system is expected to exhibit long-range, AFM
order, in agreement with experiment. 
\end{abstract}
\maketitle

\section{Introduction}

Materials based on the delafossite structure with the general chemical
formula ABO$_{2}$ exhibit interesting physical properties \citep{Shannon1971,Kawazoe1997,Wawrzynska2008,Poienar2009,Maignan2009}.
In general, \textbf{A} and \textbf{B}-sites in ABO$_{2}$ represent
monovalent and trivalent cations, respectively, having a linear and
octahedral environment of oxygen atoms. In this case, the \textbf{B}-site
which is responsible for magnetism in delafossite materials forms
an edge-shared triangular lattice or a honeycomb lattice when the
system crystallizes in hexagonal/rhombohedral or monoclinic space
groups, respectively.

There exist a variety of honeycomb materials \citep{Marquardt2006}
and in recent years attempts have been made to synthesize delafossite
materials with tetravalent ions at the \textbf{B}-site (honeycomb
lattice), which are known as 3$R$-delafossites. A few examples of
3$R$-delafossites are Ag(Li$_{1/3}$Ru$_{2/3}$)O$_{2}$ \citep{Kimber2010,Ramesha2011,Kumar2019},
Ag(Li$_{1/3}$Rh$_{2/3}$)O$_{2}$ \citep{Todorova2011} and Ag(Li$_{1/3}$Ir$_{2/3}$)O$_{2}$
\citep{Todorova2011}.\textbf{ }We would like to mention here that
Ag insertion into the primary structure of Li$_{2}$MO$_{3}$ (M =
Ru, Rh or Ir) results in its placement between consecutive metal layers
which essentially reduces the inter-layer connectivity and thus makes
the materials highly two-dimensional (2D) in nature.

As per some recent theoretical studies \citep{Khaliullin2013,Meetei2015,Svoboda2017}
novel ground state properties are expected for 4$d$/5$d$ materials
depending upon the variation of superexchange energy scale, Hund's
coupling ($J$$_{H}$) and spin-orbit coupling (SOC). In particular,
honeycomb lattice based 5$d$ materials are at the forefront of current
experimental and theoretical research because of the possibility of
stabilizing the Heisenberg-Kitaev Hamiltonian and a rich phase diagram
upon variation of the exchange couplings is envisaged \citep{Trebst2017}.
On the other hand, a different scenario might emerge while dealing
with 3$d$ transition metal ions, where SOC is much weaker than the
other energy scales, namely $U$ (on-site Coulomb repulsion) and $J$$_{H}$.
For instance, it has recently been shown by Wei \textit{et al.} \citep{Wei2011}
that the 2D honeycomb lattice based Affleck-Kennedy-Lieb-Tasaki (AKLT)
state with $S=3/2$ is a prototype for the physical realization of
the measurement based quantum computation (MBQC). According to Maciej
\textit{et al}. \citep{Koch-Janusz2015} the multiorbital insulators
in the framework of Hubbard models with nearest-neighbor hopping on
a honeycomb lattice could even lead to the $S=3/2$ AKLT state. In
light of the aforementioned proposals, it is worth investigating the
physical properties of a honeycomb lattice with $S=3/2$, with the
intention of exploring some novel ground state properties.

Herein, we report for the first time the sample preparation, structural,
and physical properties of a 3$d$ transition metal oxide (delafossite)
Ag(Li$_{1/3}$Mn$_{2/3}$)O$_{2}$. In this compound, the Mn$^{4+}$
ions ($S=3/2$) form a 2D honeycomb lattice with Li-ions positioned
at the center of each honeycomb unit. The material was structurally
characterized by a combination of x-ray and neutron diffraction measurements,
along with magnetization, specific heat, and electron spin resonance
(ESR). In addition, $^{7}$Li nuclear magnetic resonance (NMR) spectra
and spin-lattice relaxation rate measurements were performed. Structural
characterization done with x-ray and neutron diffraction suggest a
superstructure (honeycomb) formed by magnetic Mn$^{4+}$ions in the
crystallographic $ab$-plane. Interestingly, the same asymmetric peak,
seen in the paramagnetic region, is found to get more intense on lowering
the temperature, as observed in our neutron diffraction studies. Magnetization
data show an anomaly in the vicinity of 50~K and antiferromagnetic
coupling is inferred from our susceptibility analysis. Another thermodynamic
measurement, specific heat, also confirms this anomaly and locates
the magnetic ordering at 47~K. The integrated intensities of the
electron spin resonance ESR absorption lines mimic the bulk magnetic
susceptibility. The ESR measurements evidence a critical broadening
as a function of temperature with a transition temperature $T_{{\rm N}}$
= 45 K. Further, the line broadening on approaching $T_{{\rm N}}$
may be alternatively described in terms of a Berezinski-Kosterlitz-Thouless
(BKT) scenario with $T_{{\rm KT}}=40(1)$~K. The static susceptibility
(free from defects/impurity) tracked using local probe $^{7}$Li NMR
spectra measurements also reproduce the anomaly observed in bulk susceptibility
measurements, while the $^{7}$Li spin-lattice relaxation rate does
not show any sharp anomaly around the 50 K transition. This is likely
due to the symmetric location of the $^{7}$Li with respect to the
magnetic Mn$^{4+}$ ions, giving rise to a filtering of antiferromagnetic
fluctuations. Our experimental results corroborate the establishment
of a magnetically ordered ground state for the 3$d$-based system
Ag(Li$_{1/3}$Mn$_{2/3}$)O$_{2}$ which is also supported by first
principles electronic structure calculation.

\section{Experimental and Computational details}

The polycrystalline samples of the quaternary 3R-delafossite oxide
Ag{[}Li$_{1/3}$Mn$_{2/3}${]}O$_{2}$ were prepared by a combination
of sol-gel and ion-exchange methods. First, the starting material
Li$_{2}$MnO$_{3}$ was prepared \textcolor{black}{by sol-gel route
and then fired at 500}°\textcolor{black}{C for 6 hours. After confirming
by x-ray diffraction that }Li$_{2}$MnO$_{3}$ was\textcolor{black}{{}
single phase, AgNO$_{3}$}\textcolor{red}{{} }\textcolor{black}{was
mixed with }Li$_{2}$MnO$_{3}$\textcolor{black}{{} in the ratio 1 :
3. The resultant mixture was slowly heated to 300}°\textcolor{black}{C
and held for 6 hours following which the desired material }Ag{[}Li$_{1/3}$Mn$_{2/3}${]}O$_{2}$\textcolor{black}{{}
was obtained by removing the residual byproduct LiNO$_{3}$ by washing
the mixture with water. }

\textcolor{black}{The room temperature powder x-ray diffraction (xrd) measurements}
were performed using a PANalytical Xpert Pro x-ray diffractometer
with Cu-K$_{\alpha}$ radiation ($\lambda$= 1.5418\,$\textrm{Å}$).
Neutron diffraction data were taken on the DMC beamline at the Paul
Scherrer Institute PSI using a wavelength $\lambda=2.4586$ Å. Microstructural
investigations were performed with a CM 200 Philips transmission electron
microscope (TEM) operating at 200\,kV. Magnetization $M$ measurements
in the temperature range 2-400~K as a function of applied field $H$
were performed on a Quantum Design SQUID VSM with the powder sample
loaded in a capsule and for measurements in the 400-630\,K range,
the high temperature option of the Quantum Design VSM was used. The
heat capacity measurements were carried out on a Quantum design PPMS
using the thermal-relaxation method. ESR was measured in an ELEXSYS
E500 spectrometer (Bruker) at X-band frequency of 9.4 GHz in a magnetic
field of about $H$ = 18 kOe. The spectrometer was equipped with a
helium gas-flow cryostat ESR 900 (Oxford Instruments) operating in
the temperature range $T=$ 4 - 300 K. The polycrystalline samples
were immersed in paraffin in suprasil quartz glass tubes and mounted
in the cavity. ESR detects the microwave absorption due to magnetic
dipolar transitions induced between the Zeeman levels of the sample
as a function of the external magnetic field. Due to the lock-in amplification
technique by field modulation in ESR, one records the field derivative
of the absorption spectra. The $^{7}$Li-NMR measurements (spectra
and spin-lattice relaxation rate 1/$T_{1}$) were performed both in
the fixed field (93.9543\,kOe) and swept field ($\nu=95$\,MHz)
mode to gain further insights into the intrinsic static susceptibility
of Mn-moments in Ag$_{3}$LiMn$_{2}$O$_{6}$ by measuring the line
shift as a function of temperature. 

All the electronic structure calculations based on DFT presented in
this paper were carried out in the planewave basis within generalized
gradient approximation (GGA)\textbf{ }\citep{Perdew1996} of the Perdew-Burke-Ernzerhof
exchange correlation supplemented with Hubbard U as encoded in the
Vienna\textbf{ ${\it ab}$-${\it initio}$ }simulation package (VASP)
\citep{Kresse1993,Kresse1996} with projector augmented wave potentials
\citep{Blochl1994,Kresse1999}. The calculations are done with usual
values of U and Hund's coupling (\textbf{$\mathrm{\mathrm{J}}_{\mathrm{H}}$)
}chosen for Mn with U$_{eff}$ ~($\equiv$U- $\mathrm{\mathrm{J}}_{\mathrm{H}}$)
= 3.0~eV in the Dudarev scheme \citep{Dudarev1998}. In order to
achieve convergence of energy eigenvalues, the kinetic energy cut
off of the plane wave basis was\textbf{ }chosen to be 600 eV. The
Brillouin-Zone integrations are performed with 8 $\times$ 4 $\times$
6 Monkhorst grid of $k$-points. The exchange paths were identified
by calculating hopping parameters by constructing the Wannier function
using the VASP2WANNIER and the WANNIER90 codes \citep{Mostofi2014}.
In addition, to get the\textbf{ }minimum energy structure, symmetry
protected ionic relaxation was been carried out using the conjugate-gradient
algorithm until Hellman-Feynman forces on each atom were less than
the tolerance value of 0.01 eV/Å.\textbf{ }

\section{results and discussion}

We will now present the results of our various bulk and local probe
measurements on Ag$_{3}$LiMn$_{2}$O$_{6}$.

\subsection{Structure analysis }

\begin{figure}
\begin{centering}
\includegraphics[scale=0.35]{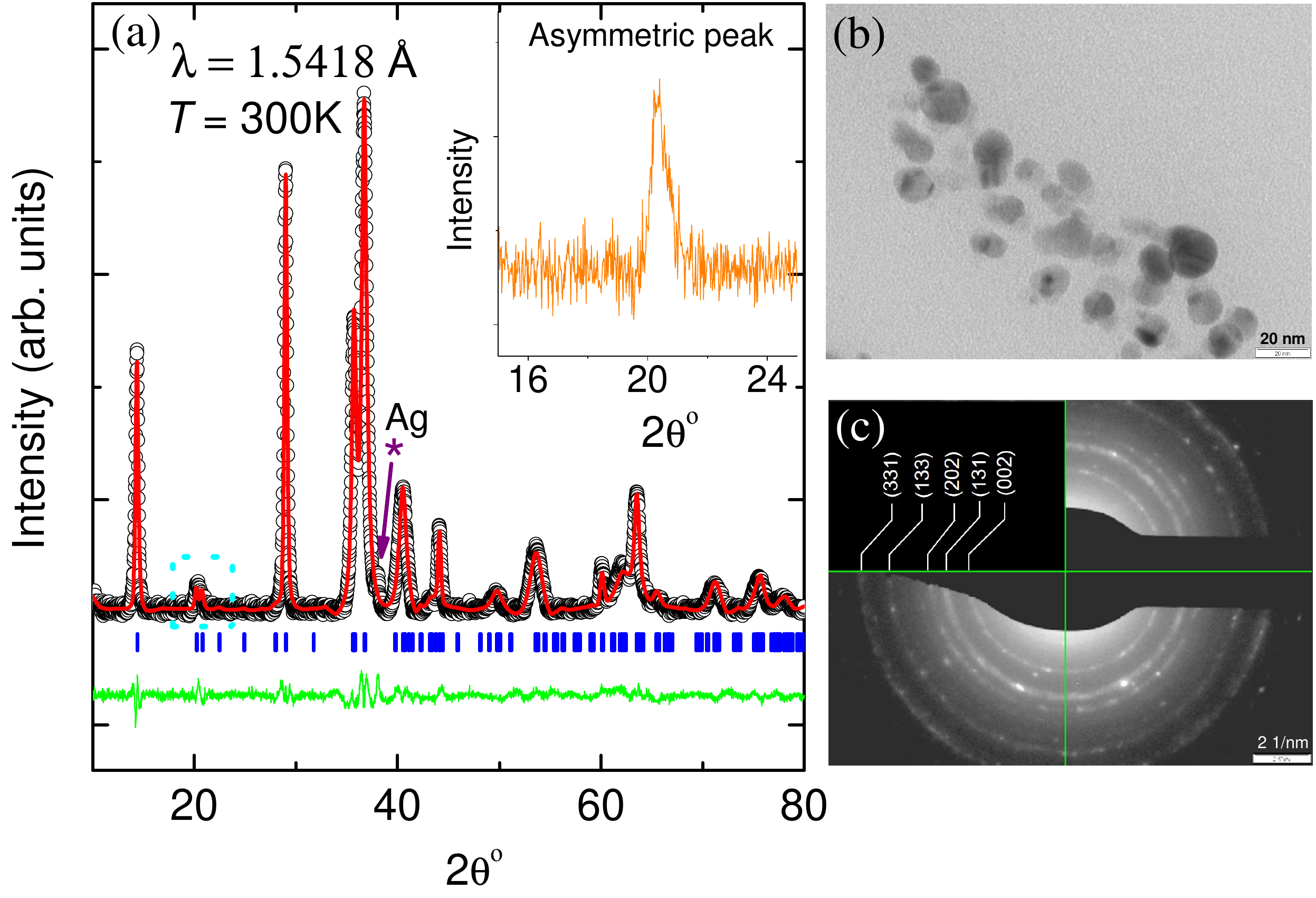} 
\par\end{centering}
\caption{\label{Ag3LiMn2O6 XRD 1.5}  (a) X-ray diffraction pattern
for Ag$_{3}$LiMn$_{2}$O$_{6}$ with $\lambda=1.5418\,\textrm{Å}$
at 300\,K. Experimental data (black open circles) and Rietveld refined
data (red solid line) are shown. The Bragg positions and the difference
pattern are shown by blue vertical lines and green line, respectively.
The asymmetric peak seen around 21°~is marked with a dotted rectangular
box and inset shows the magnified version of the portion in the box.
The peak marked with an asterik ({*}) is an impurity peak due to unwashed
Ag remaining in the final product. (b) TEM image of Ag$_{3}$LiMn$_{2}$O$_{6}$
nanoparticles. (c) Selected Area Electron Diffraction (SAED) of Ag$_{3}$LiMn$_{2}$O$_{6}$
nano particles where the (hkl) indexing of various rings is shown.}
\end{figure}

Figure~\ref{Ag3LiMn2O6 XRD 1.5} depicts the x-ray diffraction
pattern for Ag$_{3}$LiMn$_{2}$O$_{6}$ at 300\,K. The x-ray diffraction
pattern for Ag$_{3}$LiMn$_{2}$O$_{6}$ was found to be similar to
the isostructural material Ag$_{3}$LiRu$_{2}$O$_{6}$ \citep{Kimber2010}
of the quaternary 3R-delafossite family. A peak corresponding to a
small amount of Ag is seen in the xrd pattern. An asymmetric
reflection (a superstructure peak) was also seen around 2$\theta\approx$
21° (see the inset of Fig. \ref{Ag3LiMn2O6 XRD 1.5}). This asymmetric
peak (more prominent in neutron diffraction), commonly known as the
Warren peak \citep{Warren1941}, arises as a consequence of the irregular
stacking sequences of layers in a structure. In the present case,
Mn$^{4+}$ ions are found to form 2D honeycomb layers in the $ab$-plane
and the irregular stacking pattern results from the stacking faults
which then limit the correlation length in the crystallographic $c$-direction.
The x-ray diffraction peaks for Ag$_{3}$LiMn$_{2}$O$_{6}$ were
found to be broader than those of Ag$_{3}$LiRu$_{2}$O$_{6}$ and
the particle size determined from the Scherrer formula is estimated
to be about 20~nm. We then performed microstructure analysis from
the TEM and selected area electron diffraction (SAED) images depicted
in Fig. \ref{Ag3LiMn2O6 XRD 1.5}. SAED analysis shows that the continuous
ring patterns (see inset of \ref{Ag3LiMn2O6 XRD 1.5}(c)) from our
polycrystalline sample could be well indexed with the hexagonal lattice
of Ag$_{3}$LiMn$_{2}$O$_{6}$. Our TEM analysis showed that the
mean size of nanoparticles was between 10-20 nm and the nanoparticles
appeared to be nearly spherical in morphology. This value is in good
agreement with results obtained from XRD. The low-temperature synthesis
route of the starting material Li$_{2}$MnO$_{3}$ (see experimental
section) could possibly be the reason for the nano crystallinity of
this material.

\begin{figure}
\begin{centering}
\includegraphics[scale=0.05]{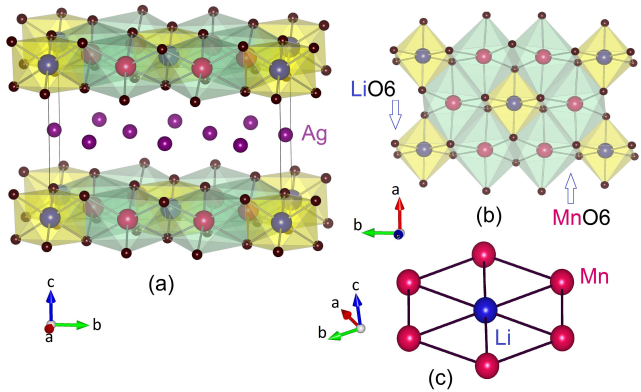} 
\par\end{centering}
\caption{ \label{Crystal structure} (a) Unit cell of Ag$_{3}$LiMn$_{2}$O$_{6}$
in a monoclinic symmetry with space group $C\,2/m$. (b) Formation
of edge-sharing LiO$_{6}/$MnO$_{6}$ octahedra in $ab$-plane. (c)
2D honeycomb lattice of Mn atoms having Li-atom at its center. }
\end{figure}

In order to extract information about the unit cell parameters and
atomic positions, the x-ray diffraction pattern of Ag$_{3}$LiMn$_{2}$O$_{6}$
was refined under the FullProf Suite program \citep{Carvajal1990}
using the structural parameters of Ag$_{3}$LiRu$_{2}$O$_{6}$ as
an initial model. All the Bragg reflections obtained for Ag$_{3}$LiMn$_{2}$O$_{6}$
could be successfully indexed with a monoclinic space group $C\,2/m$
and the refined atomic coordinates and lattice constants are listed
in Table \ref{atomic coordinates}. Because of the particle size being
in the nanometer range, evident from the broadened x-ray peaks and
TEM images, microstructure parameters were also taken into account
while refining the crystal structure and a quadratic form of strain
formation under Laue class \textcolor{black}{$mmm$ was considered.
}The Rietveld refinement quality factors expressed by $R_{wp}$, $R_{exp}$,
$R_{p}$ and $\chi^{2}$ have the values 2.89\%, 1.82\%, 2.22\% and
2.52, respectively. The crystal structure of Ag$_{3}$LiMn$_{2}$O$_{6}$
based on x-ray diffraction refinement is shown in Fig.~\ref{Crystal structure}.
The MnO$_{6}$ octahedra form an edge-sharing, 2D, honeycomb network
in the crystallographic $ab$-plane and the LiO$_{6}$ octahdera sit
at the center of the honeycomb network, see Figs. \ref{Crystal structure}(b)
and (c). Intercalated Ag atoms go in-between the consecutive honeycomb
layers.\textbf{ }

\begin{figure}
\begin{centering}
\includegraphics[scale=0.37]{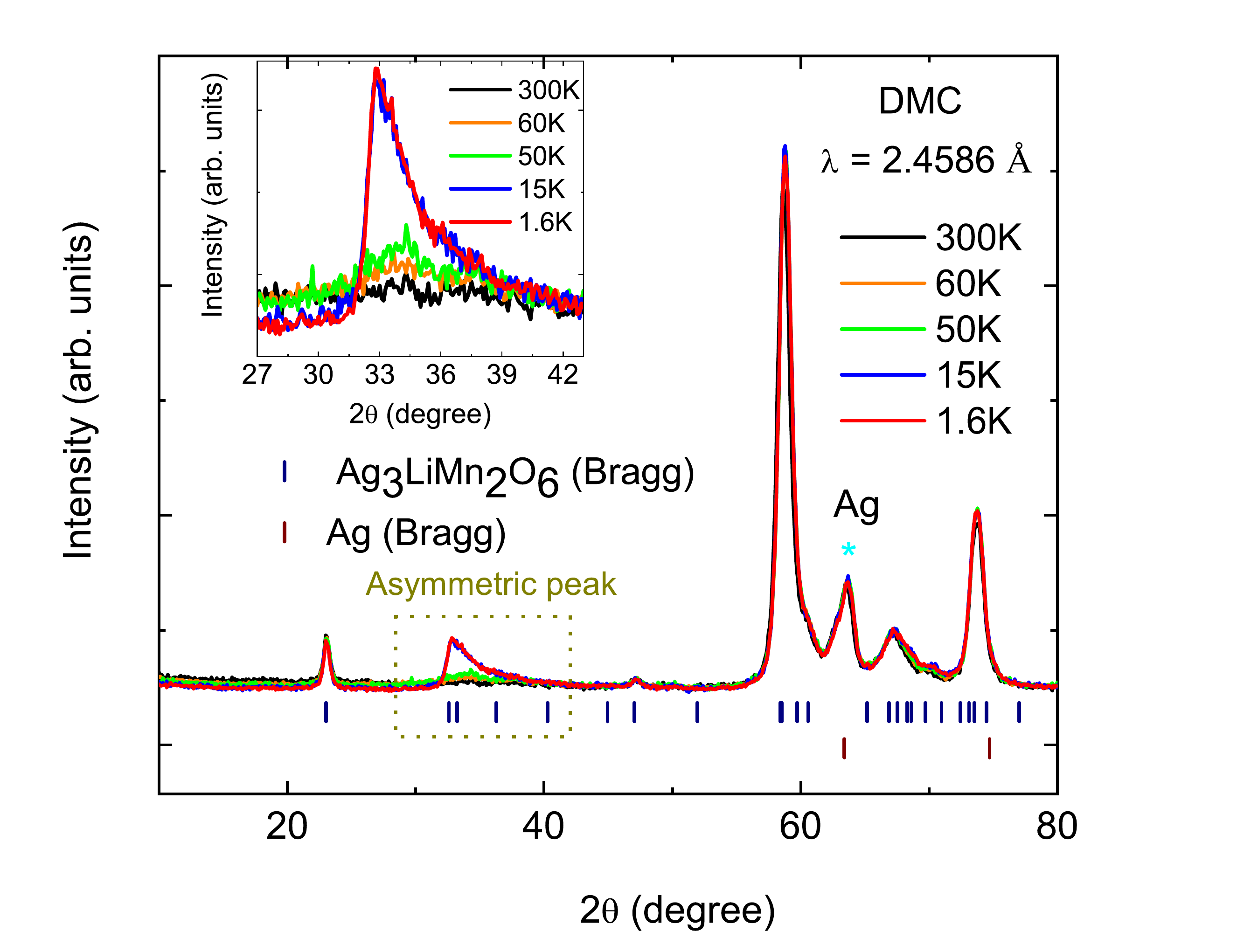} 
\par\end{centering}
\caption{\label{ Mn-neutron}  Neutron diffraction data collected
for Ag$_{3}$LiMn$_{2}$O$_{6}$ with $\lambda=2.4586\,\textrm{Å}$
at different temperatures. Vertical green and black lines represent
the Bragg positions for Ag$_{3}$LiMn$_{2}$O$_{6}$ and the impurity
phase Ag, respectively. An enhancement in the intensity of peak at
$33$°~is due to the appearance of magnetic order. The impurity peak
of Ag is marked with an asterisk.}
\end{figure}

\begin{table*}
\begin{centering}
\caption{\label{atomic coordinates} The structural parameters obtained after
Rietveld refinement of the x-ray diffraction data collected at a wavelength
of \textgreek{l} = 1.5418 Å for Ag$_{3}$LiMn$_{2}$O$_{6}$ under
the space group $C\,2/m$ at 300\,K. The obtained lattice constants
are $a=5.0690(7)\,$$\textrm{Å}$, $b=8.7863(11)\,$$\textrm{Å}$
and $c=6.3755(9)\,$$\textrm{Å}$ with $\beta=74.938(11)^{\circ}$.}
\par\end{centering}
\medskip{}

\centering{}%
\begin{tabular}{l>{\centering}p{1cm}>{\centering}p{1.8cm}>{\centering}p{1.8cm}>{\centering}p{1.8cm}>{\centering}p{0.9cm}>{\centering}p{0.8cm}}
\hline 
Atoms  & site  & \textcolor{black}{x}  & y  & z  & B$_{iso}$  & Occ\tabularnewline
\hline 
Ag1  & 2d  & 0  & 0.5  & 0.5  & 0.52  & 1\tabularnewline
Ag2  & 4h  & 0.5  & 0.3271(4)  & 0.5  & 0.52  & 1\tabularnewline
Li  & 2a  & 0  & 0  & 0  & 0.49  & 1\tabularnewline
Mn  & 4g  & 0  & 0.6648(11)  & 0  & 0.45  & 1\tabularnewline
O1  & 8j  & 0.4270(23)  & 0.3471(17)  & 0.8420(18)  & 0.36  & 1\tabularnewline
O2  & 4i  & 0.1459(45)  & 0.5  & \centering{}0.1393(36)  & 0.36  & 1\tabularnewline
\hline 
\end{tabular}
\end{table*}
In our neutron diffraction data (see the inset of Fig. \ref{ Mn-neutron}),
an asymmetric peak is seen to emerge below about 50 K while all the
other peaks are nearly unchanged. This must be from the ordering transition
which is evident in other measurements such as magnetic susceptibility,
heat capacity, etc. which are detailed in the following sections.

\subsection{Magnetization }

\begin{figure}
\begin{centering}
\includegraphics[scale=0.35]{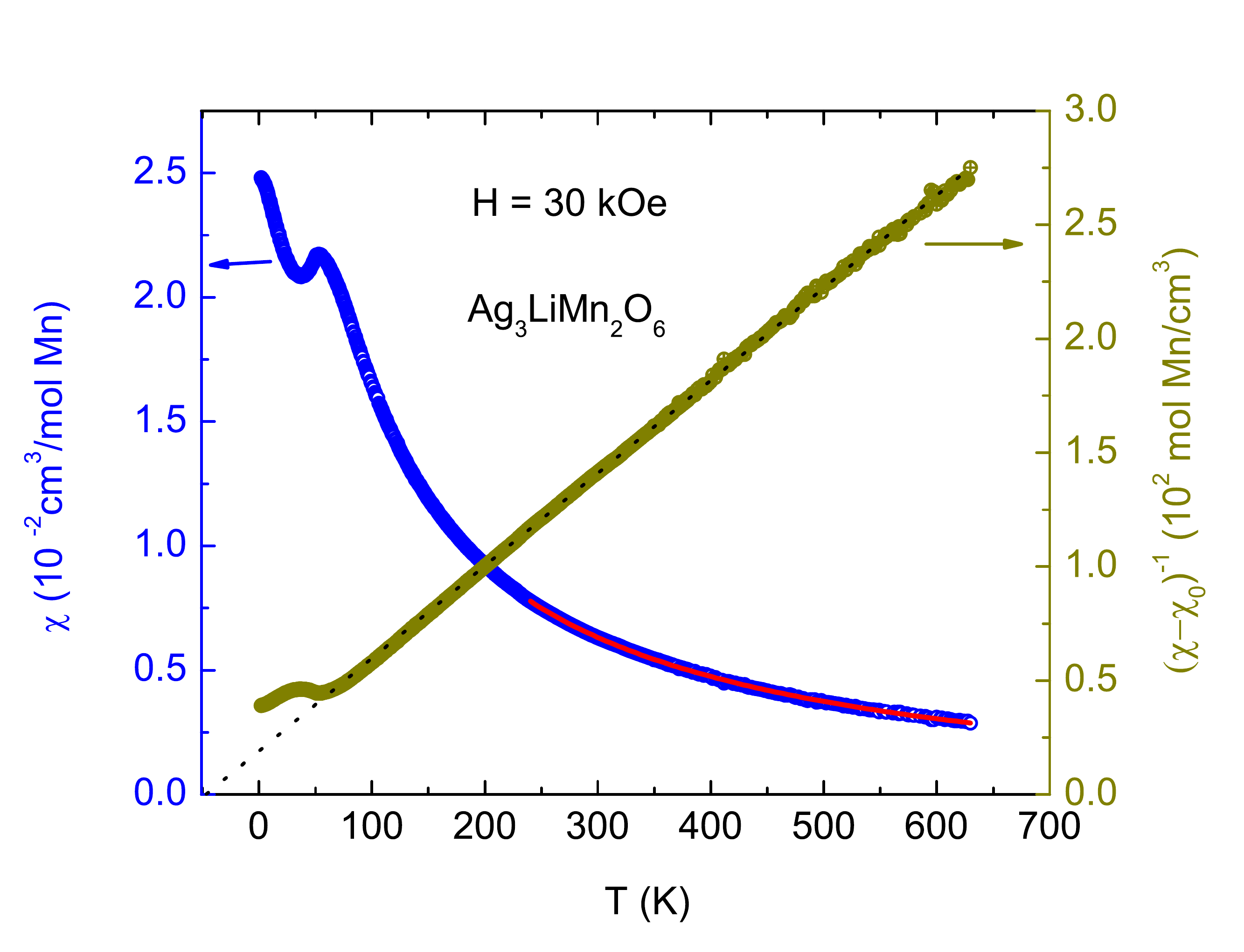} 
\par\end{centering}
\caption{\label{Mn magnetization}  Left $y$-axis: variation
of the susceptibility in the temperature range \textcolor{black}{$2-630$}\,K
with an applied field of 30~kOe and the Curie-Weiss fit (red line)
in the temperature range \textcolor{black}{$240-630$}\,K. Right
$y$-axis: inverse susceptibility, free from temperature independent
susceptibility ($\chi_{0}$), in the temperature range \textcolor{black}{$2-630$}\,K.
The intercept (dotted black line) on the $x$-axis gives a Curie-Weiss
temperature of about $-51\,$K.}
\end{figure}

Figure \ref{Mn magnetization} depicts the dc susceptibility ($M$/$H$)
of Ag$_{3}$LiMn$_{2}$O$_{6}$ measured in the temperature range
\textcolor{black}{$2-630$}\,K with an applied field of 30~kOe.
The susceptibility shows a gradual increase on lowering the temperature
before exhibiting a well rounded anomaly at around 50\,K following
which it exhibits an upturn. The anomaly observed at 50\,K might
be a signature of magnetic order, while the low-$T$ increase of susceptibility
could be partly due to some extrinsic contributions and/or defects
(see ESR/NMR results later in the paper). The Curie-Weiss fit ($\chi=\chi_{0}+C/(T-\theta_{CW})$)
to the susceptibility data in the temperature range \textcolor{black}{$240-630$}\,K
yields: temperature independent susceptibility $\chi_{0}=-7.835\times10^{-4}$\,cm$^{3}$/mol
Mn, the Curie-Weiss temperature $\theta_{CW}$ $\sim$ -51(1)\,K,
indicative of antiferromagnetic interactions among Mn$^{4+}$ moments,
and a Curie constant $C=2.495\pm0.014$~cm$^{3}$ K/mole Mn or an
effective paramagnetic moment$\sim4.46\pm0.33\,\mu_{B}$. The electronic
configuration of Mn$^{4+}$ is 3$d$$^{3}$ and hence the expected
effective moment (considering $g=2$ as obtained from electron spin
resonance discussed later) is 3.87 $\mu_{B}$. Note that the value of
$\chi_{0}$ can not be determined very accurately as the susceptibility
has still not leveled off even above 600 K. Consequently, there is
some uncertainty in the determination of the Curie constant and $\theta_{CW}$. Further, 10-20 nm size grains are present in our sample. Defects
on the surfaces of nanoparticles could also stabilize a moment and
contribute to the observed value.  A higher than expected effective moment was also observed in CaMnO$_3$ which increased with oxygen depletion. \citep{Zeng1999}

\subsection{Specific heat }

\begin{figure}
\begin{centering}
\includegraphics[scale=0.35]{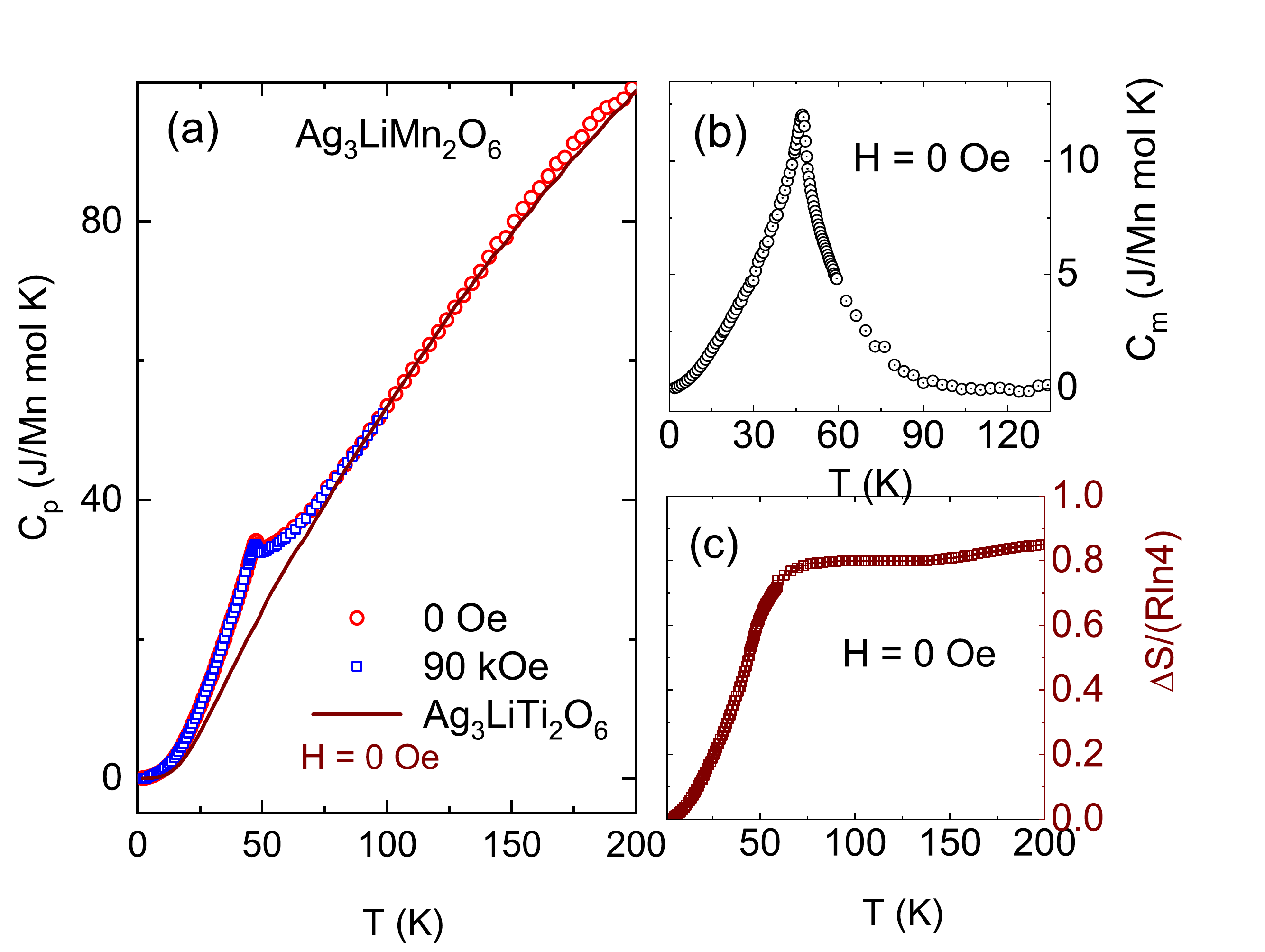} 
\par\end{centering}
\caption{\label{Ag3LiMn2O6 Cp(T) data}  (a) Specific heat of
Ag$_{3}$LiMn$_{2}$O$_{6}$ measured with $H=0$~Oe and $90$~kOe
and nonmagnetic Ag$_{3}$LiTi$_{2}$O$_{6}$ (solid line: brown) at
$H$ $=0$~Oe , used for the estimation of lattice contribution.
(b) The magnetic specific heat, $C_{m}$, of Ag$_{3}$LiMn$_{2}$O$_{6}$
at $H$ $=$ $0$\,Oe. (c) Entropy change obtained at $H=0$~Oe
normalised to Rln(4). }
\end{figure}

The specific heat of the cold pressed powder sample of Ag$_{3}$LiMn$_{2}$O$_{6}$
was measured and an anomaly was also noticed there as was observed
in our magnetization data. Figure~\ref{Ag3LiMn2O6 Cp(T) data}(a)
depicts the temperature variation of specific heat $C_{p}(T)$ of
Ag$_{3}$LiMn$_{2}$O$_{6}$ in the temperature range \textcolor{black}{$1.8-200$}\,K
in 0 and 90 kOe magnetic field. The $C_{p}(T)$ data distinctly show
the presence of an anomaly around $48$\,K, which is found to be
insensitive to the applied magnetic field. \textcolor{black}{The magnetic
}specific heat \textcolor{black}{contribution }($C_{m}$)\textcolor{black}{{}
in }Ag$_{3}$LiMn$_{2}$O$_{6}$\textcolor{black}{{} was obtained by
subtracting the lattice }specific heat\textcolor{black}{{} $C_{lattice}$.
To estimate the $C_{lattice}$ of }Ag$_{3}$LiMn$_{2}$O$_{6}$\textcolor{black}{,
an isostructural material Ag$_{3}$LiTi$_{2}$O$_{6}$ was prepared
and its }specific-heat\textcolor{black}{{} was measured in zero magnetic
field and prior to subtracting this from the }$C_{p}(T)$ data\textcolor{black}{{}
of }Ag$_{3}$LiMn$_{2}$O$_{6}$\textcolor{black}{{} it was scaled with
the }$C_{p}(T)$ data\textcolor{black}{{} of }Ag$_{3}$LiMn$_{2}$O$_{6}$\textcolor{black}{.}
Initially, Bouvier scaling \textcolor{black}{\citep{Bouvier1991}}
was used to scale the lattice specific heat of\textcolor{black}{{} Ag$_{3}$LiTi$_{2}$O$_{6}$},
which gives a correction factor ($\frac{\theta_{D}(Mn)}{\theta_{D}(Ti)}$
= 0.988) to the temperature axis, where $\theta_{D}(Mn)$ and $\theta_{D}(Ti)$
are the Debye temperatures for the Mn and Ti compounds, respectively.
However, this does not appear to reliably estimate the lattice specific
heat of Ag$_{3}$LiMn$_{2}$O$_{6}$. In fact, it was found to exceed
the total specific heat of Ag$_{3}$LiMn$_{2}$O$_{6}$. We then manually
matched the specific-heat of nonmagnetic sample, in the high temperature
region, by rescaling its temperature axis by multiplying it by 1.085.
The magnetic specific heat $C_{m}$ is plotted in Fig.~\ref{Ag3LiMn2O6 Cp(T) data}(b).
\textcolor{black}{The }$C_{m}$ data clearly show a $\lambda$-like
anomaly at 47\,K. The experimental magnetic entropy change $\Delta S$,
estimated from the $C_{m}-T$ curve, is about 80$\pm5$\% of the theoretical
value of 11.52 J/Mn mol K for spin $S=3/2$, as can be seen in Figure~\ref{Ag3LiMn2O6 Cp(T) data}(c).
The transition seen at 47~K approximately accounts for 55\% entropy
change while nearly the rest of the entropy is recovered above 70~K.
The recovery of a significant amount of entropy above the ordering
temperature is probably suggestive of the presence of short-range
magnetic correlations. One should recall that the intensity of asymmetric
peak seen in neutron diffraction measurements (see inset of Fig. \ref{ Mn-neutron})
also does not immediately collapse to the peak recorded at 300~K,
indicating again that magnetic correlations survive at least up to
60~K and are in-line with our specific heat analysis.

\subsection{ESR Results}

\begin{figure}
\begin{centering}
\includegraphics[scale=0.25]{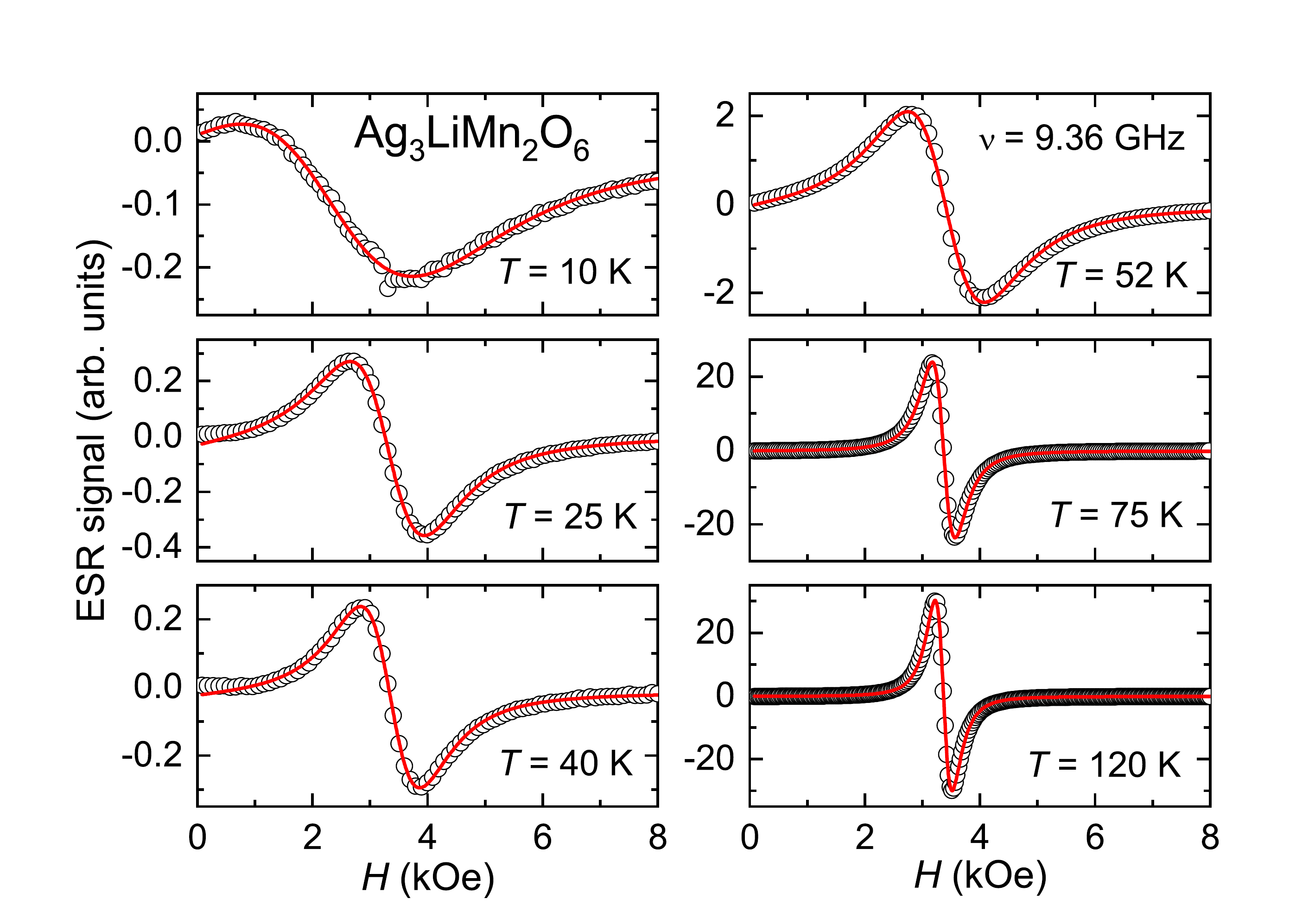} 
\par\end{centering}
\caption{\label{Fig. 6}  ESR spectra of Ag$_{3}$LiMn$_{2}$O$_{6}$
obtained at X-band frequency for different temperatures. Solid lines
indicate Lorentz curves as described in the text.}
\end{figure}

To investigate the correlated magnetism originating as a result of
interacting Mn-moments, arranged in a honeycomb geometry, ESR measurements
were performed as a function of temperature. In order to study the
spin dynamics of Ag$_{3}$LiMn$_{2}$O$_{6}$ and to get the evolution
of the corresponding ESR parameters with temperature, the ESR line
shape was analyzed. Typical ESR spectra of a powder sample of Ag$_{3}$LiMn$_{2}$O$_{6}$
are shown in Fig. \ref{Fig. 6}. All spectra consist of a single exchange-narrowed
resonance line --- i.e., any line splitting or inhomogeneous broadening
is averaged out by the isotropic exchange interaction. The resonance
line is well described by a Lorentz shape at resonance field $H_{{\rm res}}$
with half width at half maximum $\Delta H$ including the counter
resonance at $-H_{{\rm res}}$ and a small contribution of dispersion
(D) to the absorption (A) given by the (D/A) ratio in case of large
line width as described in Ref. \citep{Joshi2004}. The results of
ESR line shape fitting for some representative spectra are shown by solid
lines in Fig.~\ref{Fig. 6}.

\begin{figure}
\begin{centering}
\includegraphics[scale=0.3]{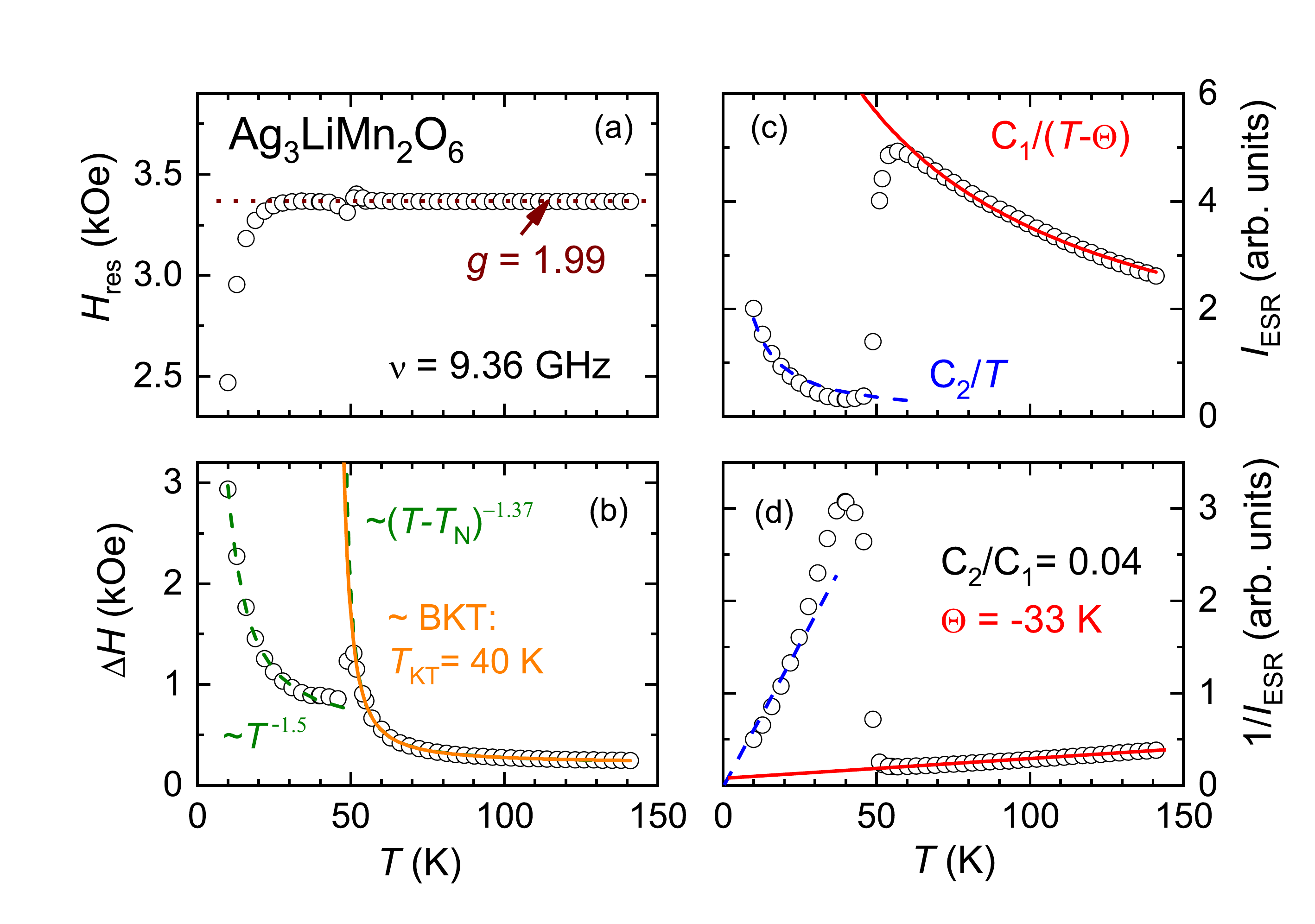} 
\par\end{centering}
\caption{\label{Fig. 7}  The temperature dependence of the ESR
parameters of Ag$_{3}$LiMn$_{2}$O$_{6}$ (a) resonance field (dashed
line indicates $g=1.99$), (b) linewidth with critical behavior (dashed)
and BKT scenario (solid), (c) ESR intensity, and (d) inverse intensity
with Curie-Weiss law (solid) and low-temperature Curie law (dashed).}
\end{figure}
The ESR line is related to the manganese exchange coupled system.
The temperature dependencies of the resonance field, the linewidth,
the intensity, and the inverse intensity are summarized in Fig. \ref{Fig. 7}.
Starting with the resonance field (see Fig. \ref{Fig. 7} (a))in the
upper left frame, we obtain a $g$ value of $g=h\nu/\mu_{B}H_{{\rm res}}=1.99$
at high temperature which is typical for Mn$^{4+}$ ions (spin $S=3/2$)
in an octahedral ligand field with its half-filled $t_{2g}$ triplet.
On decreasing temperature, the $g$-factor remains approximately constant
down to a weak anomaly at $T_{{\rm N}}=48$~K and then decreases
on further cooling followed by a divergence on approaching zero temperature.

Focusing on the integrated intensity and its inverse representation
shown in the frames on the right hand side in Fig.~\ref{Fig. 7},
the data above $T_{{\rm N}}$ are perfectly described by a Curie-Weiss
law $I=C_{1}/(T-\theta)$ with a Weiss temperature $\theta=-33$~K.
Below $T_{{\rm N}}$ the temperature dependence of the intensity is
well approximated by a pure Curie law $I=C_{2}/T$ (dashed line).
As the integrated intensity corresponds to the spin susceptibility,
it turns out from the comparison of the prefactors $C_{1}$ and $C_{2}$
that below $T_{{\rm N}}$ only 4\% of the spins contribute to the
ESR signal, i.e. weakly interacting spins, which are not involved
in the antiferromagnetic order e.g. at the surface of the powder grains
or at defect sites.

Turning finally to Fig.~\ref{Fig. 7} (b), the linewidth is found
to depend strongly on temperature, indicating two different spin-dynamic
regimes above and below $T_{{\rm N}}$. Starting from a value of about
200 Oe at high temperature, the linewidth is found to increase with
decreasing temperature, diverging close to the antiferromagnetic phase
transition with values larger than 1 kOe. On further cooling, the
linewidth recovers to about 800 Oe but diverges again with values
above 3 kOe below 4 K.

The broadening of the ESR line on approaching $T_{{\rm N}}$ may be
treated in terms of critical behavior due to slowing down of spin
fluctuations in the vicinity of an order-disorder transition. This
causes the divergence of the spin correlation length, which in turn
affects the spin-spin relaxation time of exchange narrowed ESR lines
resulting in the critical broadening, given by 
\begin{equation}
\Delta H(T)=\Delta H_{0}+A\left(\frac{T_{{\rm N}}}{T-T_{{\rm N}}}\right)^{\beta}
\end{equation}
where the first term $\Delta H_{0}$ describes the limiting temperature-independent
linewidth for the exchange narrowed regime, while the second term
reflects the critical behavior with the critical exponent $\beta$.
The dashed and the solid lines on the left lower panel in Fig.~\ref{Fig. 7}
represent a least-squares-fit of the linewidth data. The best fitting
was obtained with the parameters $\Delta H_{0}=221(2)$~Oe, $A=13(2)$~kOe,
$T_{N}=45(1)$~K and $\beta=1.37(5)$. The value of $T_{{\rm N}}$
obtained here is close to that from our heat capacity data. Below
$T_{{\rm N}}$, the residual ESR signal diverges approximately with
a power law $T^{-1.5}$.

Given the fact that the critical exponent of the divergence above
$T_{{\rm N}}$ is significantly smaller than the expected value of
2.6 for the 2D Heisenberg antiferromagnetic model\citep{Benner1990},
the line broadening on approaching $T_{{\rm N}}$ may be alternatively
described in terms of a Berezinski-Kosterlitz-Thouless (BKT) scenario
like in the case of the honeycomb system BaNi$_{2}$V$_{2}$O$_{8}$
{[}Ref. \citep{Heinrich2003}{]} with spin $S=1$ Ni$^{2+}$. Indeed
the temperature dependence of the linewidth is very well approximated
by the expression 
\begin{equation}
\Delta H=A\exp(3b/(T/T_{{\rm KT}}-1)^{0.5})+\Delta H_{0}
\end{equation}
with the Kosterlitz-Thouless temperature $T_{{\rm KT}}=40(1)$~K,
the parameter $b=0.83(6)$, the prefactor $A=11(2)$~Oe and the residual
linewidth $\Delta H_{0}=196(5)$~Oe, as shown in the lower left frame
of Fig.~\ref{Fig. 7} The BKT scenario indicates the spin-spin relaxation
via magnetic vortices, governed by the vortex correlation length which
diverges at $T_{{\rm KT}}$ due to vortex-antivortex pairing. Originally
this topological phase transition was derived for the XY model by
Berezinskii \citep{Berezinskii1971} and by Kosterlitz and Thouless
\citep{Kosterlitz1973}. But later on it was shown \citep{Cuccoli2003}
that already a weak anisotropy is enough to provide a BKT scenario.
Concerning 2D spin $S=3/2$ antiferromagnets a BKT type scenario was
reported for the triangular lattice antiferromagnets $A$CrO$_{2}$
with $A=$ H, Li, Na, Cu Ag, Pd \citep{Hemmida2011, Alexander2007}. In those Heisenberg
antiferromagnets the frustration of the antiferromagnetic couplings
gives rise to so called $Z_{2}$ vortices \citep{Kawamura1984}. Returning
to the present Mn system, the obtained fit parameters are comparable
to those found in BaNi$_{2}$V$_{2}$O$_{8}$. The parameter $b<\pi/2$
is in the range of theoretically sound values. The Kosterlitz-Thouless
temperature of about $0.83T_{{\rm N}}$ is typical for quasi-2D-antiferromagnets,
where the 3D antiferromagnetic order masks the Kosterlitz-Thouless
transition. Using the relation\citep{Bramwell1993} 
\begin{equation}
\frac{T_{{\rm N}}-T_{{\rm KT}}}{T_{{\rm KT}}}=\frac{4b^{2}}{[\ln(J/J^{\prime})]^{2}}
\end{equation}
derived for quasi 2D antiferromagnets (exchange constant $J$) with
weak planar anisotropy and inter-plane coupling $J^{\prime}$ we obtain
a ratio $J/J^{\prime}\approx40$, if we use the experimental value
of $b$, (or $\approx800$, if we use $b=\pi/2$).  Appearance of LRO close to $\theta_{CW}$ in spite of the 2D nature might result from a renormalisation of $T_{\rm N}$ due to a large in-plane AF correlation length. \citep{Chakravarty1988}

\subsection{NMR Results }

In order to develop a microscopic understanding of the magnetic properties,
$^{7}$Li ($I$ $=3/2$, $\frac{\gamma}{2\pi}=16.546$\,MHz/T) nuclear
magnetic resonance (NMR) measurements were carried out on polycrystalline
Ag$_{3}$LiMn$_{2}$O$_{6}$.

\begin{figure}
\includegraphics[scale=0.3]{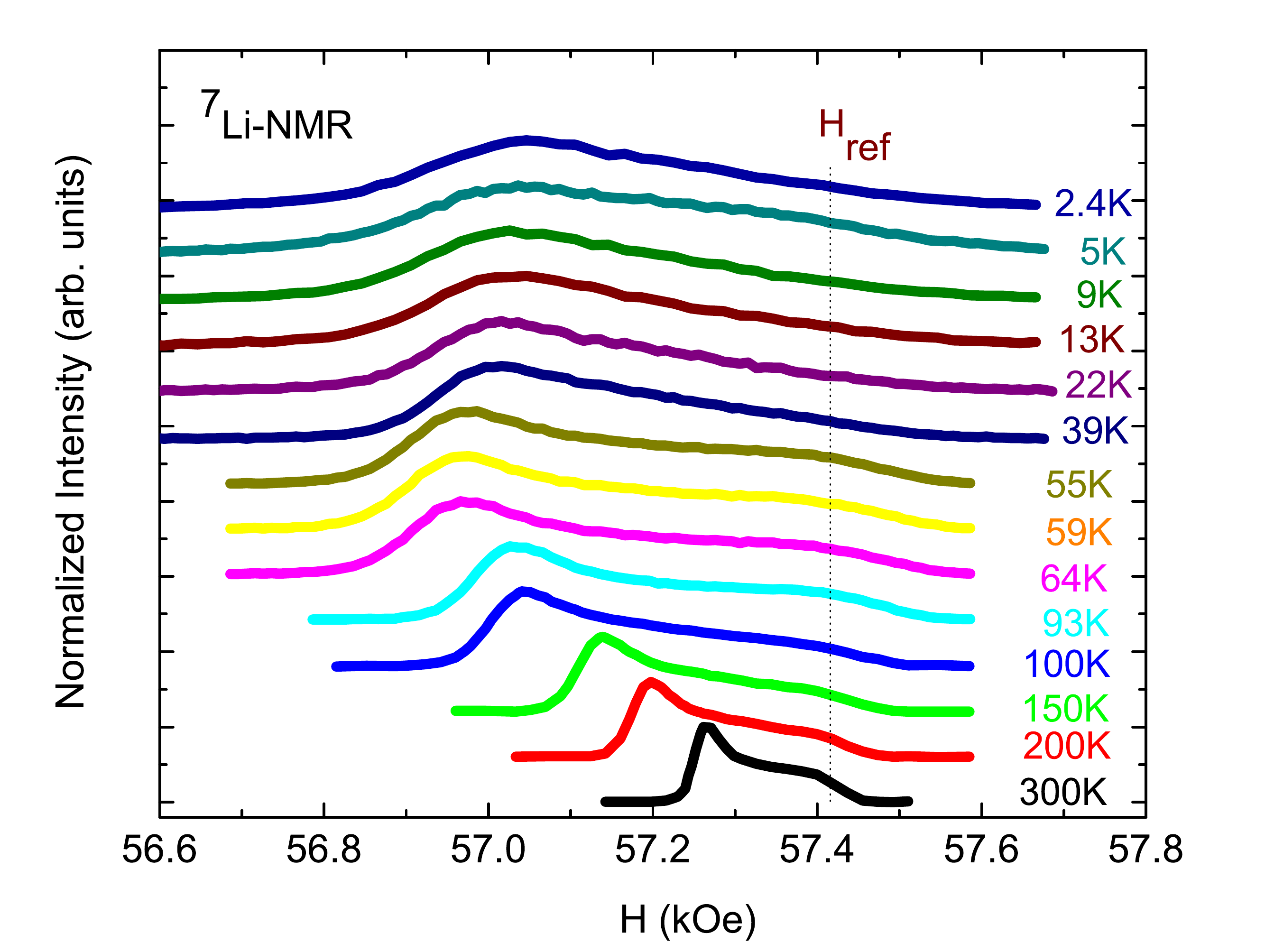}\caption{\label{Fig. 8} $^{7}$Li NMR spectra measured at different
temperatures with fixed field 93.54~kOe ($T$-range 300-80~K) and
by sweeping the field at the transmitter frequency 95~MHz ($T$-range
\textcolor{black}{$2.4-93$}\,K). The swept field spectra were corrected
for an average offset field of about 14\,Oe because of\textcolor{black}{{}
a small offset in the field of the field sweep magnet compared to
the fixed field magnet}. The fixed field data were scaled to the sweep
field data using an appropriate multiplier. The vertical dotted line
indicates the $^{7}$Li-NMR reference field measured for an isostructural
diamagnetic sample Ag$_{3}$LiTi$_{2}$O$_{6}$.}
\end{figure}

The obtained $^{7}$Li-NMR spectra in the entire measured temperature
range display a shoulder along with the main line, see Fig. \ref{Fig. 8}.
The main line is found to be broadened and shifted to the lower field
side as a function of temperature with respect to the $^{7}$Li-NMR
line measured for the isostructural diamagnetic sample Ag$_{3}$LiTi$_{2}$O$_{6}$,
while the shoulder remained almost unshifted on lowering the temperature.
The asymmetry (shoulder) in spectra could result from the anisotropy of hyperfine coupling\textcolor{black}{.} It must
be noticed that, this asymmetry is most likely not from any chemical
disorder between Li/Ru as our x-ray diffraction refinement results
discard this possibility. Surprisingly, the asymmetry in spectra was
also seen in the isostructural material Ag$_{3}$LiRu$_{2}$O$_{6}$
{[}Ref. \citep{Kumar2019}{]}. Interestingly, the hydrogenated analogue
of Ag$_{3}$LiIr$_{2}$O$_{6}$, i.e. H$_{3}$LiIr$_{2}$O$_{6}$
does not show significant asymmetry in $^{1}$H NMR line shape \citep{Kitagawa2018}.

\begin{figure}

\includegraphics[scale=0.3]{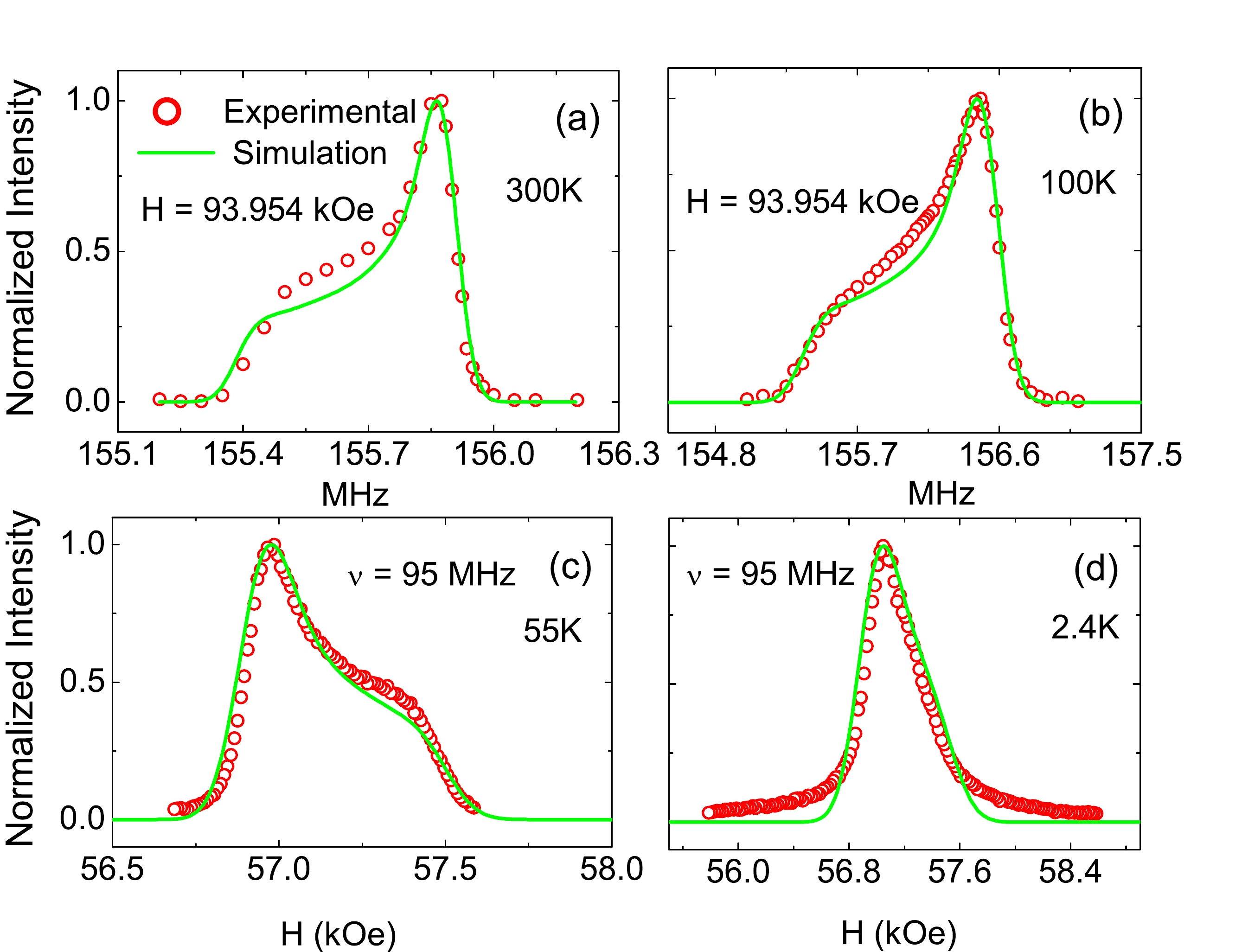}\caption{\label{Fig. 9} Simulated $^{7}$Li NMR spectra and
experimentally collected $^{7}$Li NMR spectra (for Ag$_{3}$LiMn$_{2}$O$_{6}$)
at different temperatures are shown by green solid lines and red open
circles, respectively for fixed field data ((a) and (b)) and field
sweep data ((c) and (d)).}

\end{figure}

\begin{figure}

\includegraphics[bb=0bp 0bp 800bp 590bp,clip,scale=0.3]{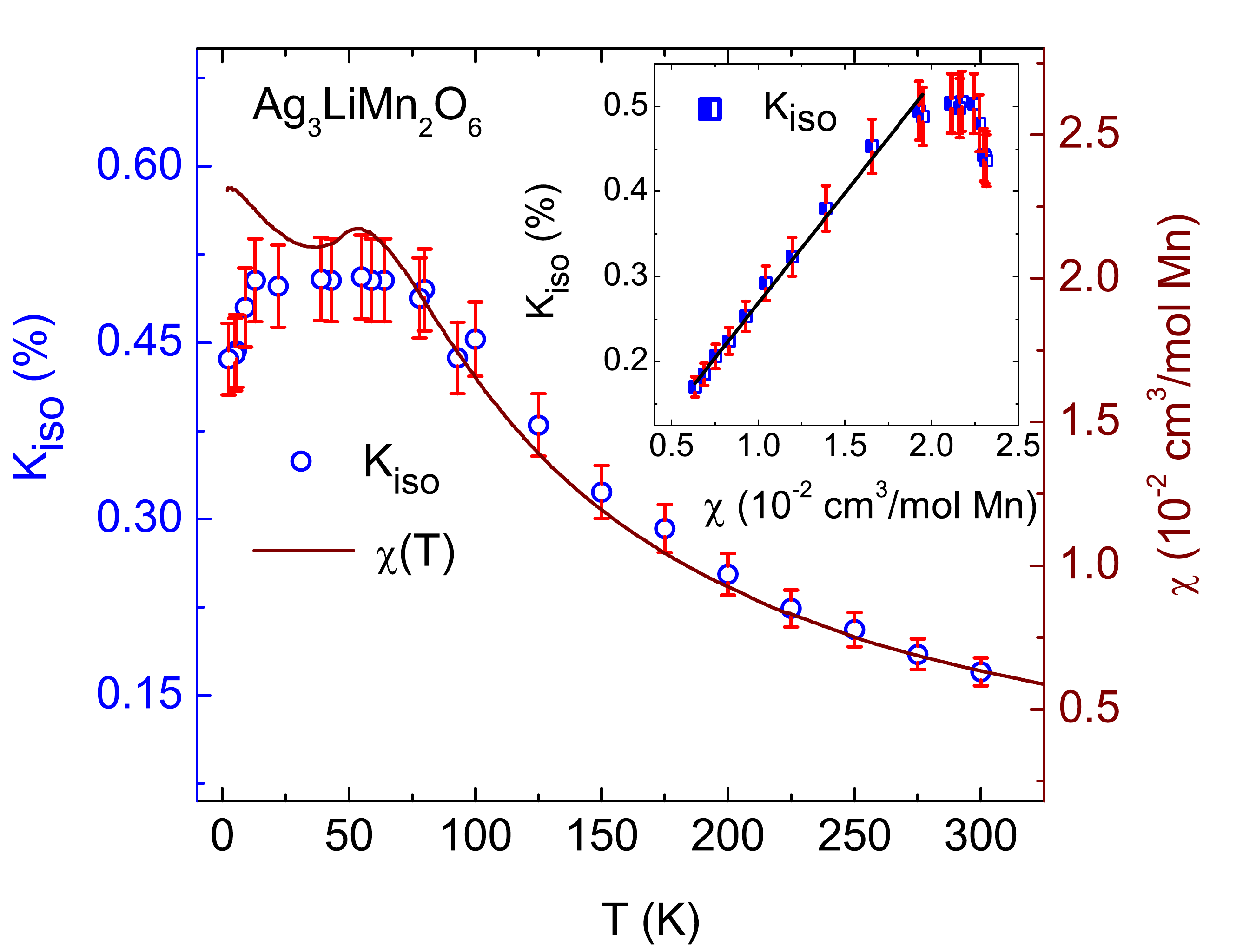}\caption{\label{Fig. 10} The left $y$-axis shows the $K_{iso}$
(blue open circles) as a function of temperature and the right $y$-axis
depicts the bulk susceptibility (solid line) data for Ag$_{3}$LiMn$_{2}$O$_{6}$.
The inset shows a plot of $K_{iso}$ versus $\chi$ with temperature
as an implicit parameter.}
\end{figure}

We analyzed the powder averaged $^{7}$Li NMR spectra of Ag$_{3}$LiMn$_{2}$O$_{6}$,
similar to the previously studied material Ag$_{3}$LiRu$_{2}$O$_{6}$,
by fitting the spectra to a combination of the anisotropic shift parameters
$K_{iso}$ and $K_{aniso}$ . A few representative simulated patterns
are shown in Fig.~\ref{Fig. 9}. The extracted $K_{iso}$ from $^{7}$Li
NMR spectra in the temperature range \textcolor{black}{$300-3$}\,K
as a function of temperature is shown in Fig.~\ref{Fig. 10}. The
$K_{iso}$ data follow the bulk susceptibility data down to about
50 K and show an anomaly around the ordering temperature. The low-$T$
deviation of the bulk susceptibility from the NMR shift could be the
result of some extrinsic impurity contribution or intrinsic defects.
The hyperfine coupling constant ($A_{hf}$) and the chemical shift
obtained from the $K$-$\chi$ plot (inset of Fig.~\ref{Fig. 10})
yield 1.45 $\pm$ 0.04\,kOe/$\mu_{B}$ and 0.009(6)\%, respectively.

The $^{7}$Li NMR spin-lattice relaxation rate ($1/T_{1}$) measurements
were performed with the intention to study the low-energy spin dynamics
or to probe the $q$-averaged dynamical susceptibility of Ag$_{3}$LiMn$_{2}$O$_{6}$
in the temperature range \textcolor{black}{$2-300$}\,K at the transmitter
frequency 95\,MHz. The saturation recovery method was employed to
measure the $1/T_{1}$ data. In order to deduce the spin-lattice relaxation
time from the measured data, the data were fitted to a combination
of two exponential decays given by the equation:

\[
(1-m(t)/m(0))=Aexp(-x/T_{1L})+Bexp(-x/T{}_{1S})
\]

where $T_{1L}$ and $T_{1S}$ are the long and short components of
spin-lattice relaxation time with $A$ and $B$ being constants.

\begin{figure}
\includegraphics[scale=0.3]{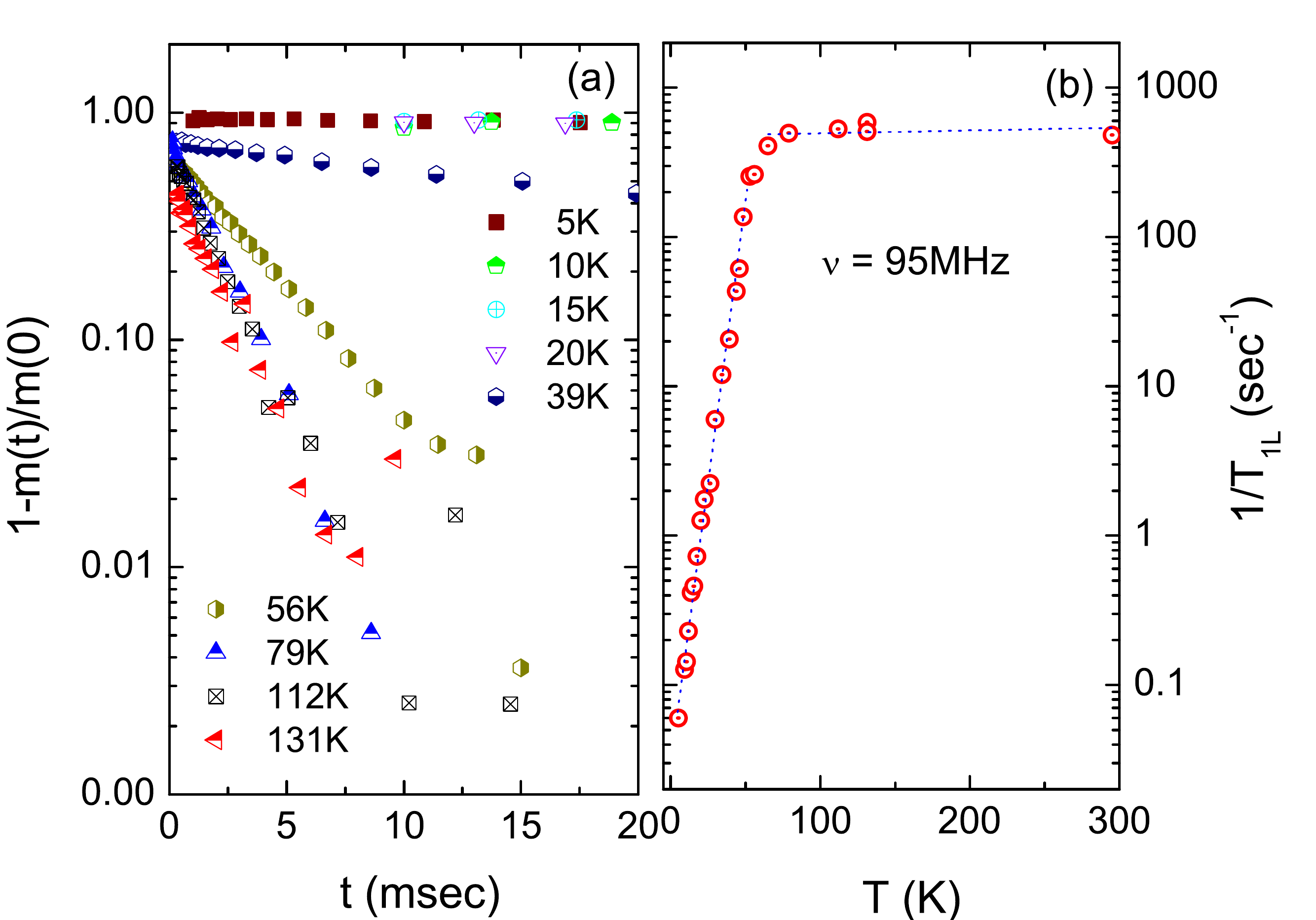}

\caption{\label{Fig. 11} (a) Recovery of the $^{7}$Li longitudinal
nuclear magnetisation as a function of delay time at a few representative
temperatures. (b) Spin-lattice relaxation rate as a function of temperature
for the long component $T_{1L}$ (smaller rate).}

\end{figure}

Figure~\ref{Fig. 11}(a) depicts a plot for $^{7}$Li nuclear magnetization
saturation recoveries at selected temperatures for Ag$_{3}$LiMn$_{2}$O$_{6}$.
\textcolor{black}{A two-component spin-lattice relaxation was also
observed in }Ag$_{3}$LiRu$_{2}$O$_{6}$. This could result from
an incomplete saturation of the NMR line leading to spectral diffusion
and an initial fast recovery. The long component of the spin-lattice
relaxation rate ($1/T_{1L}$) for Ag$_{3}$LiMn$_{2}$O$_{6}$ is
illustrated in Fig.~\ref{Fig. 11}(b). The $1/T_{1L}$ data in the
temperature range \textcolor{black}{$300-60$}\,K do not show any
variation as a function of temperature, however, below about 60\,K
$1/T_{1L}$ starts to deviate from this behavior. At the antiferromagnetic
ordering temperature, the $1/T_{1L}$ data should have also exhibited
a distinct anomaly, but no peak was seen and a smooth decrease was
noticed in $1/T_{1L}$ data in the temperature range \textcolor{black}{$2-50$}\,K.
The absence of the signature of any AFM order in the $1/T_{1L}$ data
most likely results from a cancellation of the antiferromagnetic fluctuations
at the Li position because of its symmetric position in a honeycomb
network of Mn atoms. So, because of the 2D symmetric arrangement of
Mn atoms around Li atom, which sits in the middle of honeycomb lattice,
the $1/T_{1L}$ data measured for $^{7}$Li nuclei will not sense
fluctuations in the hyperfine field perpendicular to the applied field.
Thus one, in principle, does not expect to see any sharp kink or any
critical divergence in the $1/T_{1L}$ data. This finding further
strengthens the idea of a symmetric crystallographic position of Li
atom, 2a (0, 0, 0), with respect to its surrounding Mn atoms in a
unit cell and also an absence of Li/Mn site disorder. The $1/T_{1L}$
becomes very long with decrease in temperature as static order develops
and there is absence of any fluctuations.


\subsection{Electronic structure calculation}

In order to identify the dominant exchange paths and the relevant
spin Hamiltonian of the system we have performed first principles
electronic structure calculation using VASP. The MnO$_{6}$ octahedral
units that host magnetism form an edge shared honeycomb geometry in
the $a-b$ plane with Li ions at the center of the honeycomb. Upon
relaxation, the distances of three nearest neighbor (NN) Mn atoms
become 2.926 Å~and 2.932 Å~with co-ordination two and one respectively.
The Mn-O-Mn angles in the respective NN paths are 102.33$^{\circ}$
and 102.61$^{\circ}$. The MnO$_{6}$ octahedra have a monoclinic
distortion and hence the Mn-O bond-lengths are unequal (1.876 Å~to
1.878 Å) as also the Mn-O-Mn bond angles (77.68$^{\circ}$ to 95.05$^{\circ}$). 

In order to understand the basic electronic structure, we have first
carried out non-spin polarized calculations enforcing the spin degeneracy.
The octahedral crystal field breaks the degeneracy of 5 $d$ states
of Mn atoms into triply degenerate t$_{2g}$ and doubly degenerate
e$_{g}$ states which get further split due to monoclinic distortion
maintaining crystal field splitting between t$_{2g}$ and e$_{g}$
states to be 2.5~eV. The calculations reveal that the Mn t$_{2g}$
states are half filled consistent with the Mn$^{4+}$ ($d$$^{3}$)
configuration resulting in a metallic solution. 

\begin{figure}[!h]
\includegraphics[width=1\columnwidth]{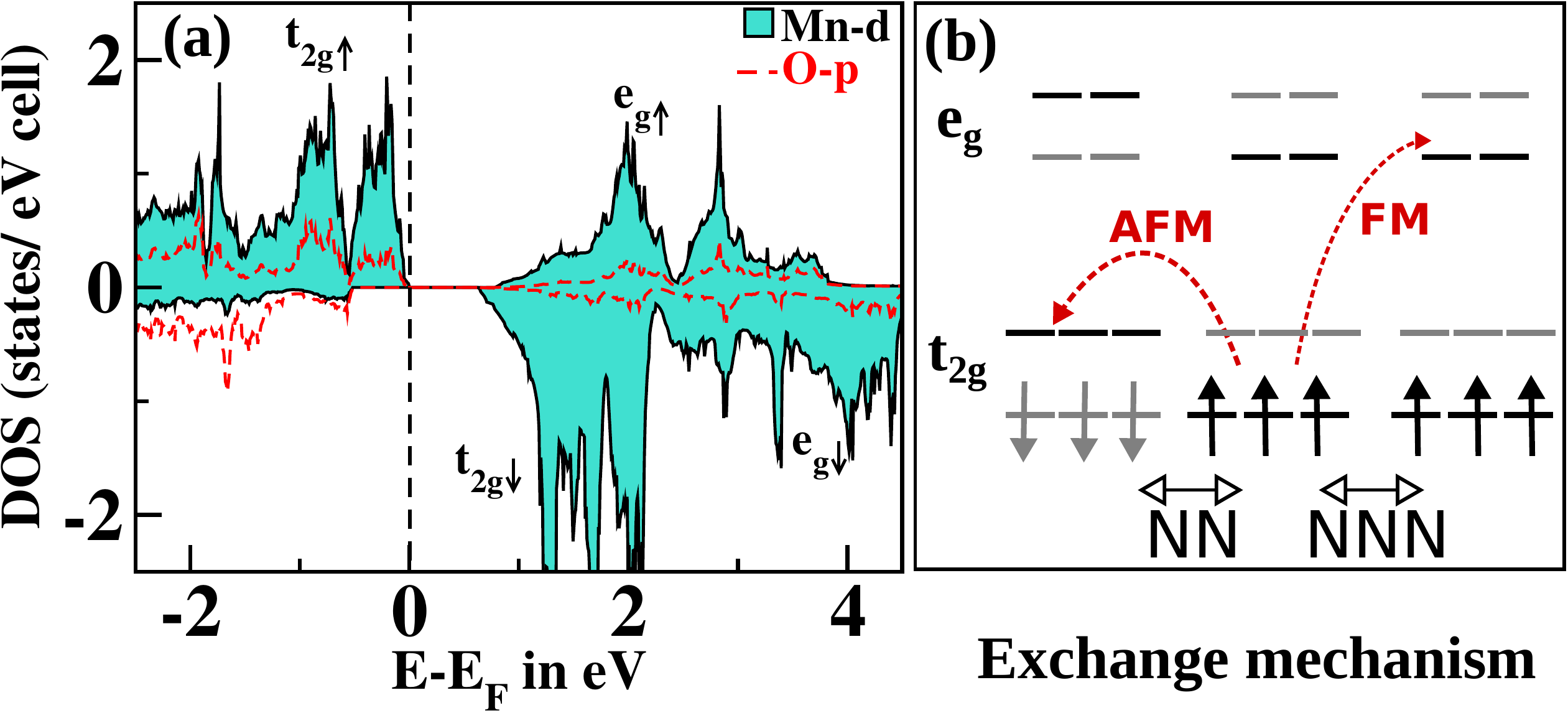} \caption{\label{magFig}  (a) The orbital decomposed spin polarized
DOS. The cyan shaded region and the red dotted lines, respectively,
show the contribution of Mn-$d$ and O-$p$ states per atom. (b) The
schematic diagram for the NN AFM and NNN FM exchange interaction mechanism.}
\end{figure}

Next we have performed spin polarized calculations with FM arrangement
of Mn spins within the GGA approximation. A plot of the spin polarized
density of states (DOS) for the FM configuration shown in Fig.\ref{magFig}(a)
reveals that in the majority spin channel the Mn-d t$_{2g}$ states
are completely filled and the minority t$_{2g}$ states are complete
empty with an exchange splitting of about 0.62~eV. The e$_{g}$ states
in both the spin channels are completely empty. In the FM calculation,
the total moment per formula unit, containing 2 Mn atoms, is calculated
to be 6.0~$\mu_{B}$ which further supports the 4$+$ charge state
of Mn and also consistent with experimentally calculated value of
$\mu_{eff}$ (4.46 $\pm$ 0.33~$\mu_{B}$). FM calculation gives
magnetic moment per Mn site to be 2.68~$\mu_{B}$ as the rest of
the moment lies in the ligand sites (0.04~$\mu_{B}$/ O) due to substantial
hybridization. Inclusion of Coulomb correlation (U) further increases
the exchange splitting (0.99 eV) and localizes the $d$ orbitals which
essentially increase the moment of Mn (2.95 $\mu_{B}$/ Mn).\\
 
In order to identify the nature of the ground state we have calculated
the energies of different possible collinear magnetic configurations
namely FM, AFM-1, AFM-2 and AFM-3. The results are shown in Table-II.
In the AFM-1 configuration, all nearest neighbor interactions (J$_{1}$
and J$_{2}$) are antiferromagnetic similar to the ground state AFM
solution of a bipartite lattice. In the AFM-2 configuration, among
the two types of NN interactions, J$_{1}$ is AFM while J$_{2}$ is
ferromagnetic. On the contrary the AFM-3 configuration has FM J$_{1}$
and AFM J$_{2}$ interactions. From Table \ref{Table:TABLE} it is
evident that AFM-1 is the lowest energy configuration with NN interactions
antiferromagnetic. 

\begin{figure}[!h]
\includegraphics[scale=0.3]{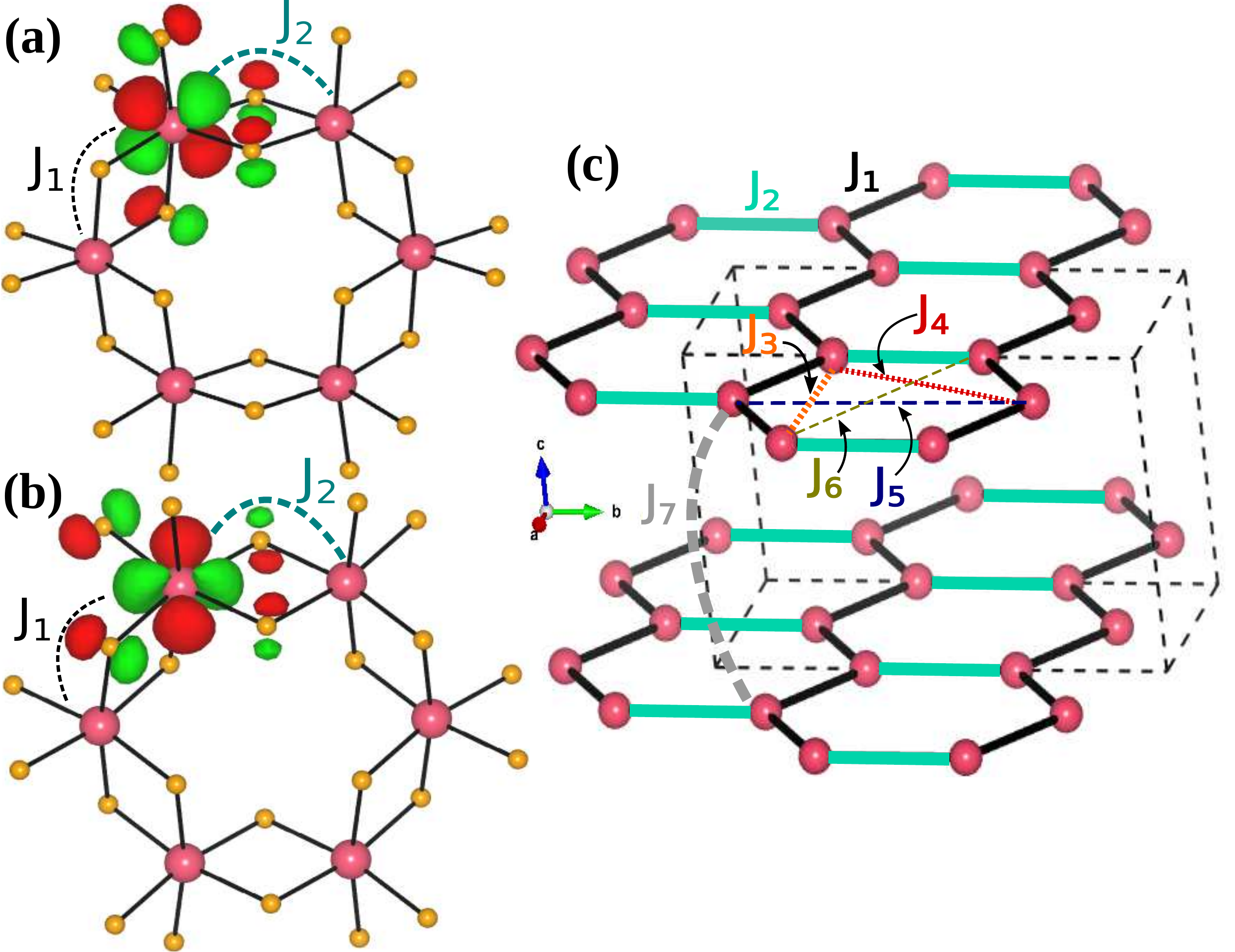} \caption{(a) and (b) are two different Wannier orbitals connecting NN Mn atoms.
(c) Different symmetric exchange interactions corresponding to neighbor
distances. NN J$_{1}$ and J$_{2}$ interactions are marked with black
and cyan solid lines. NNN interactions J$_{3}$ and J$_{4}$, 3rd
neighbors J$_{5}$ and J$_{6}$ and interplane interaction are denoted
with orange, red, brown, blue and grey dotted lines respectively.}
\label{jFig} 
\end{figure}

\begin{table}[!h]
\caption{Variation of the magnetic moments (in the GGA + U calculation) on
the various ions for different ordering arrangements and their corresponding
energies are tabulated here.}

\selectlanguage{british}%
\resizebox{0.5\textwidth}{!}{%

\selectlanguage{english}%
\begin{tabular}{cccccc}
\hline 
Magnetic\  & Total Moment\  & Moment/Mn\  & Moment/O\  & Gap\  & $\Delta$E/f.u. \tabularnewline
Configuration & /f.u. in $\mu_{B}$  & in $\mu_{B}$  & in $\mu_{B}$  & in eV  & in meV \tabularnewline
\hline 
\hline 
AFM-1  & 0.0  & 2.88  & 0.0  & 1.54  & 0.0 \tabularnewline
AFM-2  & 0.0  & 2.91  & 0,$\pm$ 0.01  & 1.33  & 36.6\tabularnewline
AFM-3  & 0.0  & 2.89  & 0,$\pm$ 0.02  & 1.47  & 24.6 \tabularnewline
FM  & 6.0  & 2.95  & -0.01  & 0.99  & 35.5 \tabularnewline
\hline 
\end{tabular}\foreignlanguage{british}{}}

\label{Table:TABLE} 
\end{table}
Now for a quantitative estimate of Mn inter-site exchange strengths,
we have calculated symmetric exchange interactions with "Four State"
method based on the total energy of the system with few collinear
spin alignments. If the magnetism in the system is fully described
by the Heisenberg Hamiltonian, the energy for such a spin pair can
be written as follows \citep{Xiang2011}: 
\begin{equation}
E=-J_{12}S_{1}\cdot S_{2}+S_{1}\cdot h_{1}+S_{2}\cdot h_{2}+E_{all}+E_{0}\label{tothamEqn}
\end{equation}
where $J_{ij}$ is the symmetric exchange coupling along bond which
connects spin pair $i$ and $j$. $h_{1}$ = $-\sum\limits _{i\neq1,2}J_{1i}S_{i}$
and $h_{2}$ = $-\sum\limits _{i\neq1,2}J_{2i}S_{i}$ and $E_{all}$
= $-\sum\limits _{i,j\neq1,2}J_{ij}S_{i}\cdot S_{i}$ and $E_{0}$
contains all other non-magnetic energy contributions. The second (third)
term in Eqn.~\ref{tothamEqn} corresponds to the coupling of the
spin 1 (2) with all other spins in the unit cell except spin 2 (1),
$E_{all}$ characterizes the exchange couplings between all spins
in the unit cell apart from spins 1 and 2. The exchange interaction
strength between site 1 and 2 obtained with total energy of four collinear
spin alignments (such as ($1_{\uparrow}$~$2_{\uparrow}$), ($1_{\uparrow}$~$2_{\downarrow}$),
($1_{\downarrow}$~$2_{\uparrow}$), ($1_{\downarrow}$~$2_{\downarrow}$))
has the expression \citep{Xiang2011} \\
 
\begin{equation}
J_{12}=-\frac{E_{\uparrow\uparrow}+E_{\downarrow\downarrow}-E_{\uparrow\downarrow}-E_{\downarrow\uparrow}}{4S^{2}}
\end{equation}
\\
 The first (second) suffix of energy ($E$) tells the spin state of
site 1 (2). The obtained symmetric exchange interactions are J$_{1}$=
-2.59 meV (AFM), J$_{2}$= -2.28 meV (AFM), J$_{3}$ = 0.54 meV (FM),
J$_{4}$= 0.69 meV (FM), J$_{5}$= -0.16 meV (AFM) and J$_{6}$= -0.15
meV (AFM) and the respective exchange paths are shown in \ref{jFig}(c).
The possible mechanism of spin conserved exchange couplings for the
NN AFM exchange interactions (J$_{1}$ and J$_{2}$) and FM second
neighbors (J$_{3}$ and J$_{4}$) are shown in Fig.\ref{magFig}(b).
For NN AFM alignments, exchange splitting and inter-site hoppings
are the key parameters while for the FM arrangement, spins have to
overcome the crystal field splitting with nearest neighbor superexchange
hopping to obey the Hund's coupling. The strongest nearest neighbor
(NN) interactions along with the finite (though small) further neighbor
interactions results in the long ranged AFM ordering in the system.
The Wannier function plots in Fig. \ref{jFig}(a) and (b) clearly
shows that the NN interactions via Mn-O-Mn superexchange path are
dominant. The substantial deviation of Mn-O-Mn angle(s) ($\sim$ 102$^{\circ}$)
from $\frac{\pi}{2}$ along paths J$_{1(2)}$ favor an antiferromagnetic
alignment in accordance with Goodenough-Kanamori-Anderson rules \citep{Anderson1950,Goodenough1955,Goodenough1958,Kanamori1959},
which is consistent with our results. To check the strength of inter-layer
coupling along $c$ direction we have fixed the static intra-plane
magnetic arrangement to AFM-1 and applied the above mentioned "Four
state" method to calculate the interlayer exchange J$_{7}$. The
interplanar coupling strength is estimated to be FM in nature with
magnitude 0.14 meV. Theoretically calculated $\theta_{CW}$ from these
exchange strengths turns out to be $-54.8$K which is very close to
experimentally obtained value, $-51$K. The ratio of inter-planar
and NN intra-planar interaction ($\frac{J_{NN}}{J_{inter}}$) is nearly
19.0 which suggests that the magnetic network in primarily of 2D in
nature. 

\section{Conclusions }

In summary, a new honeycomb material Ag$_{3}$LiMn$_{2}$O$_{6}$
has been studied using x-ray diffraction, neutron diffraction, magnetization,
specific heat, ESR and NMR measurements and first principles calculations.
An asymmetric peak, signature of superstructure, is seen in both the
x-ray and neutron diffraction and it is found to grow in intensity
below about 50 K before saturating at lower temperatures. This 
suggests magnetic ordering of the honeycomb lattice. The susceptibility
measurements carried out for Ag$_{3}$LiMn$_{2}$O$_{6}$ show an
anomaly around 50\,K and the presence of antiferromagnetic interactions
($\theta_{CW}$$\sim-51$\,K) among Mn moments. The neutron diffraction
data measured down to 2\,K clearly show the onset of magnetic ordering
below 50\,K, in agreement with the anomaly observed in the susceptibility
data of Ag$_{3}$LiMn$_{2}$O$_{6}$. The heat capacity measurements
further support long-range magnetic order in Ag$_{3}$LiMn$_{2}$O$_{6}$
by exhibiting a sharp peak in the measured specific heat around 47\,K.
The entropy change inferred from the heat capacity data suggests that
the system needs to be heated to nearly 1.6 times the ordering temperature
to recover the full entropy. This  suggests that 2D magnetic correlations
 start building up at high temperature; on cooling, a
fraction of the entropy is already lost when the sample locks into LRO around 50 K. Our local probe $^{7}$Li NMR spectra measurements
done on the powder samples, which basically measure the static susceptibility
also support  thermodynamic measurements by exhibiting a
clear anomaly in the measured $^{7}$Li NMR shift. The $^{7}$Li spin-lattice
relaxation rate, which is a measure of the $q$-averaged dynamical
susceptibility, does not show a peak as observed in other measurements
presumably because of the cancellation of antiferromagnetic fluctuations
at the center (Li-site) of the hexagon also implying that the structure
remains that of a regular honeycomb with Li sitting at the centers
of the honeycomb network. However, $^{7}$Li $1/T_{1}$ too shows
a sharp decrease of nearly four orders of magnitude below about 50
K indicating the quenching of magnetic fluctuations due to the onset
of magnetic order. Taken together, our thermodynamic and $^{7}$Li NMR shift
measurements evidence the emergence of an ordered ground in the 3$d$
honeycomb material Ag$_{3}$LiMn$_{2}$O$_{6}$. Our experimental
results are corroborated by first principles electronic structure
calculations. Our $ab$ $initio$ calculations find that the NN interactions
J$_{1(2)}$ are antiferromagnetic. The further neighbor interactions
also do not give rise to any frustration. Thus, in the case of a dominant
Heisenberg term (as might be expected in this case), and even a small
inter-planar coupling (as per our calculations) the honeycomb system
displays long-range order as seen here. The manifestation of 2D effects in Ag$_{3}$LiMn$_{2}$O$_{6}$ is
seen from the analysis of the $T$-dependence of the ESR linewidth
above the transition temperature. We obtained a Kosterlitz-Thouless
temperature of about $0.83T_{{\rm N}}$ which is typical for quasi-2D-antiferromagnets.
The weak interplanar coupling is sufficient to lock the system into
3D order which then masks the Kosterlitz-Thouless transition. 

\section{Acknowledgments}

This work is partially based on experiments performed at the Swiss
spallation neutron source SINQ, Paul Scherrer Institute, Villigen,
Switzerland. We thank Department of Science and Technology (DST),
Govt. of India for financial support through the BRICS project Helimagnets.
ID thanks DST, Govt. of India and TRC for financial support. R. Kumar
acknowledges CSIR, India and IRCC, IIT Bombay for awarding him fellowships
for the completion of this work. P. M. Ette acknowledges CSIR, INDIA
for providing financial support under CSIR-SRF Fellowship (grant no.
31/52(14)2k17). AVM would like to thank the Alexander von Humboldt
Foundation for financial support during the stay at the University
of Augsburg. The work of RE were done within the framework of fundamental
research AAAA-A18-118030690040-8 of FRC Kazan Scientific Center of
RAS. A.A.Gippius acknowledges the financial support from the RFBR
Grant No. 17-52-80036. Additionally, we kindly acknowledge support from the German Research Society (DFG) via TRR80 (Augsburg, Munich).

\bibliographystyle{apsrev4-1}
\bibliography{citation_global}

\end{document}